\documentclass[a4paper,pra,twocolumn,superscriptaddress,groupedaddress]{revtex4}
\usepackage{ae}
\usepackage{physics}
\usepackage[T1]{fontenc}
\usepackage[ansinew]{inputenc}
\usepackage{amsmath}
\usepackage{amssymb}
\usepackage[caption=false]{subfig}
\usepackage{multirow}
\usepackage{array}
\usepackage[]{graphicx}
\usepackage{wrapfig}
\usepackage{makecell}
\usepackage{epstopdf}
\usepackage{color}
\usepackage[colorlinks]{hyperref}
\usepackage{lscape}
\usepackage{amsthm}
\graphicspath{ {./images1/} }

\usepackage{csquotes}
\usepackage{upgreek}
\usepackage{mathrsfs}
\usepackage{nameref}

\begin{document}
\title{Entanglement of Assistance as a measure of multiparty entanglement}
\author{Indranil Biswas}
\email{indranilbiswas74@gmail.com}
\affiliation{Department of Applied Mathematics, University of Calcutta, 92, A.P.C. Road, Kolkata- 700009, India}
\author{Atanu Bhunia}
\email{atanu.bhunia31@gmail.com}
\affiliation{Department of Applied Mathematics, University of Calcutta, 92, A.P.C. Road, Kolkata- 700009, India}
\author{Subrata Bera}
\email{98subratabera@gmail.com}
\affiliation{Department of Applied Mathematics, University of Calcutta, 92, A.P.C. Road, Kolkata- 700009, India}
\author{Indrani Chattopadhyay}
\email{icappmath@caluniv.ac.in}
\affiliation{Department of Applied Mathematics, University of Calcutta, 92, A.P.C. Road, Kolkata- 700009, India}
\author{Debasis Sarkar}
\email{dsarkar1x@gmail.com, dsappmath@caluniv.ac.in}
\affiliation{Department of Applied Mathematics, University of Calcutta, 92, A.P.C. Road, Kolkata- 700009, India}
\begin{abstract}
Quantifying multipartite entanglement poses a significant challenge in quantum information theory, prompting recent advancements in methodologies to assess it. We introduce the notion of  \enquote{Volume of Assistance} (VoA), which calculates the geometric mean of entanglement of assistance across all potential parties. We demonstrate the feasibility of VoA for three-qubit pure states and certain classes of pure tripartite qudit states. Extending this measure to four-qubit states and general multipartite scenarios follows a similar framework. We have done a comparative analysis to illustrate VoA's distinctiveness from established entanglement measures, notably showing it serves as an upper bound for the much celebrated generalized geometric measure (GGM). Remarkably, VoA excels in distinguishing a broad class of states that elude differentiation by the recently proposed Minimum Pairwise Concurrence (MPC) measure. Finally, VoA is applied to quantify genuine entanglement in the ground states of a three-qubit Heisenberg XY model, which highlights its practical utility in quantum information processing tasks.
\end{abstract}
\date{\today}
\maketitle
\section{Introduction}
\begin{figure}[h!]
		\centering
		\includegraphics[scale=0.33]{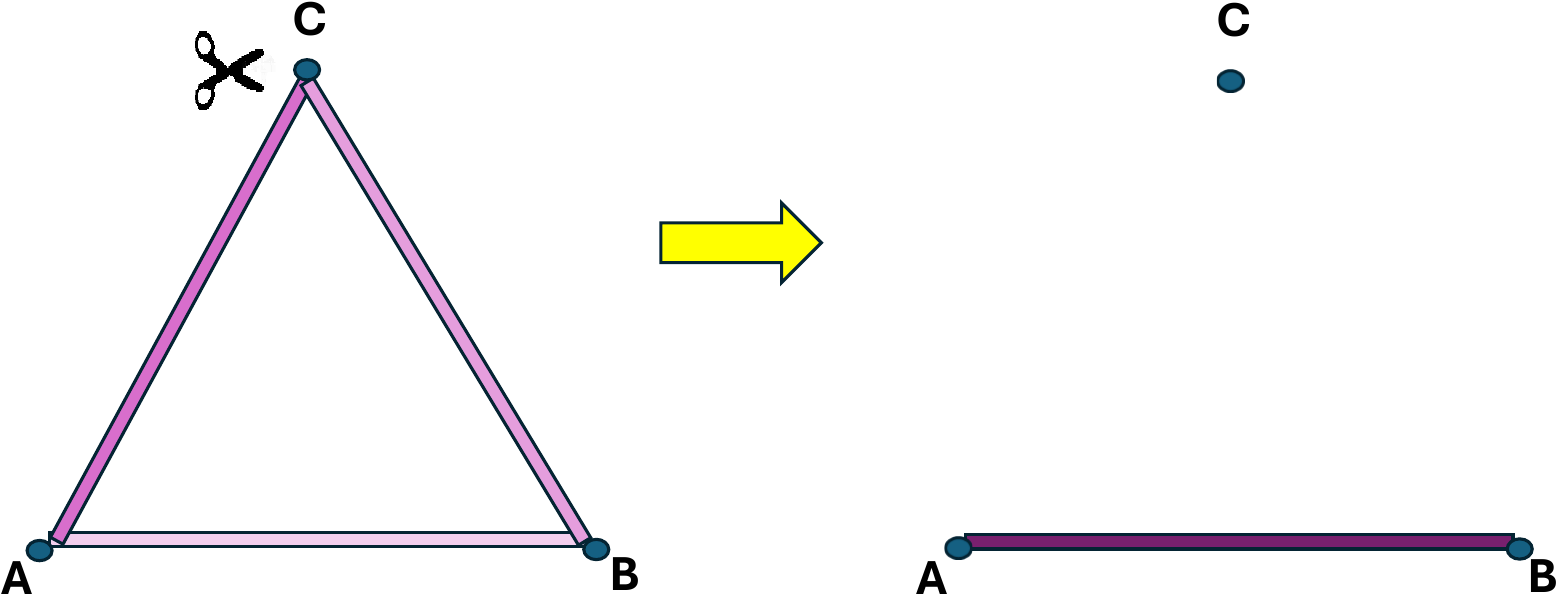}
		\caption{The scissor symbolizes the local assistance provided by Charlie ($C$) to decouple himself from the system and thereby concentrating the entanglement between Alice ($A$) and Bob ($B$).}
  \label{fig1}
	\end{figure}

The phenomenon of entanglement among quantum particles has fundamentally transformed our understanding of the universe, unveiling behaviors that were previously unimaginable within classical physics. Bell nonlocality \cite{Bell} serves as a pivotal cornerstone in both the study and characterization of entanglement. Ongoing studies are dedicated to fully harnessing quantum nonlocality, particularly in verifying its existence within network-like structures \cite{Amit1,Amit2}. Beyond the realm of entanglement, nonclassical features of quantum theory  offer vast potential. The discovery of \enquote*{Nonlocality without Entanglement} \cite{Bennett} has proven its practical utility in diverse applications such as secret sharing \cite{secret}, data hiding \cite{hiding1,hiding2,hiding3}, and quantum key distribution \cite{key}. Recent research efforts have also prioritized enhancing our comprehension and capabilities in discrimination among multipartite quantum states \cite{Halder1,Halder2,Atanu1,Atanu2,Indra,Atanu3,Atanu4,Subrata1}. These developments emphasize and also calls for developing more robust techniques to measure correlations to fully make use of the inherent nonclassicality of quantum physics.  \\
Ever since the discovery of entanglement, a formidable task has been its proper quantification. For pure bipartite states, there exist a \enquote*{good} entanglement measure, namely, Von Neumann entropy of reduced density matrices. But for generic bipartite mixed states, we lack a proper, computable entanglement measure. The situation is even worse in the case of multipartite systems, due to the existence of inequivalent SLOCC classes \cite{1}. However to harness the complete potential of entanglement, it is extremely important to precisely understand multipartite entanglement. Its proper utilisation has applications even in subjects like condensed matter physics, in particular quantum many-body systems \cite{2} apart from the uses in quantum information technologies like quantum enhanced measurements \cite{3}, quantum algorithms \cite{4}, fault-tolerant quantum computing \cite{5} and many more.\\
For such reasons, in recent times there has been a consistent effort from the researchers to come up with a proper multipartite entanglement measure. In particular, there has been a lot of effort to quantify tripartite entanglement using the characters of bipartite entanglement measures. Xie and Eberly \cite{6} have introduced triangle measure of tripartite entanglement using the entanglement polygon inequality \cite{7,8} featuring bipartite concurrence \cite{9}. Unfortunately the measure has been shown to lack the crucial property of monotonicity under LOCC \cite{10}. Apart from this, other multipartite entanglement measures has also been proposed using the complete information of reduced states \cite{11} and also by taking geometric mean of the bipartite concurrences \cite{12}. Schwaiger \textit{et. al.}, \cite{13} have also introduced an operational multipartite entanglement measure by taking an interesting approach. They have calculated the volume of states that can be converted locally to a given state and the volume of states that the given state can be converted into -- the so called source and accessible entanglement measures respectively. Another approach to find multipartite entanglement measure has been shown by taking into account thermodynamic properties of states such as local and global ergotropies \cite{14}.\\
All such measures have tried to quantify entanglement from different perspectives. However, one of the essential use of multipartite entanglement is in creating singlets at a later time. In this work we have taken this phenomena into account and introduce a novel measure of multipartite entanglement. \\
The definition of entangled multipartite systems is layered. Unlike bipartite systems, for any $n$-partite system, the entanglement can exist among any $k$ subsystems (or their all possible convex combination, for mixed states) \cite{15,16}.\\
In this paper, we are however only concerned about genuine multipartite entanglement(GME), for it constitutes the strongest form of correlation that involves all parties of the system. A good GME measure must be able to detect all such correlations and should ignore every other forms of entanglement that does not demand the same. Formally we describe this in the definition below. \\ 
\textit{Definition 1.--}
A $n$-partite GME measure $\Lambda(\cdot)$ must satisfy the following conditions \cite{17}:\\
(i) $\Lambda(\cdot)$does not increase on average under LOCC\\
(ii) for any biseparable $\rho$, $\Lambda(\rho) = 0$.\\
(iii) $\Lambda(\rho) > 0$ whenever $\rho$ is a GME.\\
As mentioned earlier, a lot of multipartite measures (ME) has been popping up for a long time. Not all of them are GME. The much celebrated tangle introduced by Coffman, Kundu and Wootters (CKW) \cite{18} fails to detect W-type states. The measures introduced in literature  \cite{19,20,21,22,23} fail to vanish for non-GME states. Nevertheless, there are some measures which do satisfy the properties of a GME measure upto some extent. In this paper we use entanglement of assistance to construct a new three qubit measure. In particular, concurrence of assistance \cite{24} (FIG. \ref{fig1}) has a computable analytical formula and forms a tripartite entanglement monotone. Later we build on this measure to create a four qubit measure.\\
The paper is organised as follows. In Sec \ref{sec2} we provide all necessary definitions related to concurrence of assistance. Sec \ref{sec3} contains the new measure for three and four qubit states along with some technical lemmas and their proofs. In Sec \ref{sec4}, we consider some examples to show the utility of our measure and show their inequivalence with some already existing measures. We provide a function to estimate the measure for three qubit mixed states in \ref{sec5}.  Finally in Sec \ref{sec6}, we conclude.\\

\section{Concurrence of Assistance}
\label{sec2}
Instead of detecting GME through construction of some functions of entanglement, another approach is to develop GME measures by observing their performance in some specific quantum information processing tasks. Recently, Choi \textit{et. al.} \cite{25} has defined a GME measure based on multi-party teleportation capability. In this paper, we have taken a somewhat similar approach by considering the localizable entanglement  (LE) \cite{26,27,28} of a GME state.
 As the name suggests, the LE between two parties Alice and Bob is the maximal amount of entanglement localized between them  by LOCC, provided the said two parties do not get involved in the process. Generally LE is very difficult to calculate. A very special case of LE is \enquote*{Entanglement of Assistance} (EoA) \cite{29,30,31,32} $\mathcal{E}_{C}^{a} (\rho _{\hspace{-0.1em}\scriptscriptstyle AB})$ defined by 
\begin{subequations}
\begin{equation}
\label{eq:1a}
    \mathcal{E}_{\scriptscriptstyle C}^{a} (\rho _{\hspace{-0.1em}\scriptscriptstyle AB}) = max \sum_i p_i \mathcal{E}(\ket{\phi_i}_{\hspace{-0.2em}\scriptscriptstyle AB})
\end{equation}
where $\mathcal{E}(\cdot)$ is a bipartite entanglement measure, $\mathcal{E}_{\scriptscriptstyle C}^{a}$ denotes the EoA with respect to the party C. The maximization is over all decompositions $\{ p_i, \ket{\phi_i}_{\hspace{-0.2em}\scriptscriptstyle AB}\}$ of the state $\rho_{AB}$ where $p_i$ denotes the probabilities satisfying $p_i\geqslant 0, \sum_i {p_i}=1$.\\
Both EoA and its generalization LE fail to become either a bipartite measure or a tripartite monotone \cite{33}. This is due to the fact that both of them do not allow Alice and Bob to perform measurements and communicate their outcomes prior to the measurement of other parties. In such cases another quantity called \enquote*{Entanglement of Collaboration} (EoC) \cite{34} proves to be a valid multipartite entanglement monotone. EoC is defined in the same way as EoA with the addition that Alice and Bob also take part in the LOCC process. However, it turns out that if we use concurrence as a measure of bipartite entanglement, then \enquote*{Concurrence of Assistance} (CoA) \cite{24} does prove to be a tripartite monotone for the pure tripartite state $\ket{\psi}_{\hspace{-0.2em}\scriptscriptstyle ABC}$ with $\rho_{\hspace{-0.1em}\scriptscriptstyle AB} = \Trace_{\scriptscriptstyle C} (\ket{\psi}\bra{\psi})$. This quantity has been shown to be a special case for which the EoA and EoC coincide \cite{35}. CoA for a tripartite state $\ket{\psi}_{\hspace{-0.2em}\scriptscriptstyle ABC}$ is defined as 
\begin{equation}
\label{eq:1b}
    \mathcal{C}_{\scriptscriptstyle C}^{a} (\ket{\psi} _{\hspace{-0.2em}\scriptscriptstyle ABC}) = max \sum_i p_i \mathcal{C}_{\scriptscriptstyle AB}(\ket{\phi_i}_{\hspace{-0.2em}\scriptscriptstyle AB})
\end{equation} 
where the maximization is over all decompositions $\{p_i, \ket{\phi_i}_{\hspace{-0.2em}\scriptscriptstyle AB}\}$  of the state $\rho_{\hspace{-0.1em}\scriptscriptstyle AB}$. $\mathcal{C}_{\scriptscriptstyle AB}(\cdot)$ denotes the function to calculate the bipartite concurrence of the state shared between parties A and B. Notice that the CoA with respect to party Charlie (C) depends solely on the decomposition of the state $\rho_{\hspace{-0.1em}\scriptscriptstyle AB}$.\\ 
Fortunately, for three qubit systems, there is a closed form analytical expression of CoA defined by 
\begin{equation}
\label{eq:1c}
    \mathcal{C}_{\scriptscriptstyle C}^{a} (\ket{\psi} _{\hspace{-0.2em}\scriptscriptstyle ABC}) = F(\rho_{\hspace{-0.1em}\scriptscriptstyle AB},\Tilde{\rho_{\hspace{-0.1em}\scriptscriptstyle AB}})
\end{equation}
where $\Tilde{\rho_{\hspace{-0.1em}\scriptscriptstyle AB}} = \sigma_y \otimes \sigma_y \rho_{\hspace{-0.1em}\scriptscriptstyle AB} \sigma_y \otimes \sigma_y$, and $F(\rho_{\hspace{-0.1em}\scriptscriptstyle AB},\Tilde{\rho_{\scriptscriptstyle AB}}) = \Trace\sqrt{\rho_{\hspace{-0.1em}\scriptscriptstyle AB}^{1/2} \Tilde{\rho_{\hspace{-0.1em}\scriptscriptstyle AB}} \rho_{\hspace{-0.1em}\scriptscriptstyle AB}^{1/2}}$ is the fidelity between $\rho_{\hspace{-0.1em}\scriptscriptstyle AB}$ and $\Tilde{\rho_{\hspace{-0.1em}\scriptscriptstyle AB}}$.
Thus it makes CoA a computable, tripartite entanglement monotone for pure tripartite states. It can also be defined for mixed states $\rho_{\hspace{-0.1em}\scriptscriptstyle ABC}$ through convex roof extension as follows:
\begin{equation}
\label{eq:1d}
    \mathcal{C}_{\scriptscriptstyle C}^{a} (\rho_{\hspace{-0.1em}\scriptscriptstyle ABC}) = min \sum_i p_i \mathcal{C}_{\scriptscriptstyle C}^{a} (\ket{\phi_i}_{\hspace{-0.2em}\scriptscriptstyle ABC})
\end{equation}
where the minimization is over all decompositions $\{p_i, \ket{\phi_i}_{\hspace{-0.2em}\scriptscriptstyle ABC}\}$ of $\rho_{\hspace{-0.1em}\scriptscriptstyle ABC}$.\\
Concurrence is only defined for two qubit systems. For a $\mathbb{C}^d\otimes \mathbb{C}^d$ systems there is a analogous, but weaker \cite{GC} quantity called \enquote*{Generalised Concurrence} (GC) \cite{35}. For a $\mathbb{C}^d\otimes \mathbb{C}^d$ state $\ket{\chi}= \sum_{i=0,j=0}^{d-1,d-1} a_{ij}\ket{ij}$, its GC is defined by
\begin{equation}
\label{eq:1e}
\mathcal{G}(\ket{\chi})= d(\lambda_0 \lambda_1 ...\lambda_{d-1})^{1/d}= d[det(A^{\dagger}A)]^{1/d}
\end{equation}
where $A=(a_{ij})$ and $(\lambda_0, \lambda_1,..., \lambda_{d-1})$ are the Schmidt numbers of $\ket{\chi}$.\\
For mixed states, it can be defined by convex roof extension
\begin{equation}
\label{eq:1f}
    \mathcal{G}(\rho)= min \sum_i p_i \mathcal{G}(\ket{\psi_i})
\end{equation}
where the minimization is taken over all decomposition of $\rho = \sum_i p_i \ket{\psi_i}\bra{\psi_i}$.\\
For a pure $\mathbb{C}^d\otimes \mathbb{C}^d\otimes \mathbb{C}^n$ tripartite pure state $\ket{\psi}_{\hspace{-0.2em}\scriptscriptstyle ABC}$, its Generalized Concurrence of Assistance (GCoA) \cite{35} with respect to C is defined by
\begin{equation}
\label{eq:1g}
     \mathcal{G}_{C}^{a} (\ket{\psi} _{\hspace{-0.2em}\scriptscriptstyle ABC}) = max \sum_i p_i \mathcal{G}(\ket{\phi_i}_{\hspace{-0.2em}\scriptscriptstyle AB})
\end{equation}
where the maximization is taken over all decompositions ${\{ p_i, \ket{\phi_i}_{\hspace{-0.2em}\scriptscriptstyle AB}\}}$ of $\rho_{\hspace{-0.1em}\scriptscriptstyle AB}=\Trace_{\scriptscriptstyle C} \ket{\psi}\bra{\psi}=\sum_i p_i \ket{\phi_i}\bra{\phi_i}$.\\
For tripartite mixed states $\rho_{\hspace{-0.1em}\scriptscriptstyle ABC}$, GCoA is defined in terms of convex roof extension:
\begin{equation}
\label{eq:1h}
     \mathcal{G}_{\scriptscriptstyle C}^{a} (\rho_{\hspace{-0.1em}\scriptscriptstyle ABC}) = min \sum_j q_j \mathcal{G}_{\scriptscriptstyle C}^{a} (\ket{\psi_j}_{\hspace{-0.2em}\scriptscriptstyle ABC})
\end{equation}
\end{subequations}
where the minimization is over all decomposition of $\rho_{\hspace{-0.1em}\scriptscriptstyle ABC}= \sum_j q_j \ket{\psi_j}\bra{\psi_j}$.\\
These are descriptions of the necessary quantities that will be used throughout the paper. Here we wrap up this section and go to the next section where we will formally introduce the measure.
\section{Volume of Assistance}
\label{sec3}
\subsection{Three qubit states}
Despite being a true tripartite monotone, either of  $\mathcal{C}_{\scriptscriptstyle C}^{a}$ or $\mathcal{G}_{\scriptscriptstyle C}^{a}$ fails to become a tripartite GME measure as it violates the condition (ii) of the Definition 1. For example, if $\ket{\psi}_{\hspace{-0.2em}\scriptscriptstyle ABC}=\frac{1}{\sqrt{2}}\ket{00+11}_{\hspace{-0.2em}\scriptscriptstyle AB}\ket{0}_{\hspace{-0.2em}\scriptscriptstyle C}$, then $\mathcal{C}_{\scriptscriptstyle C}^{a}=1$ \emph{i.e.} reaches the maximum value. However it is obvious that $\ket{\psi}_{\hspace{-0.2em}\scriptscriptstyle ABC}$ is a biseparable state. Instead, if we consider the geometric mean of the three CoAs with respect to three distinct parties, then that quantity becomes a truly three qubit GME measure and we call it the \enquote*{\emph{Volume of Assistance}} (VoA) of the tripartite state.\\
Let us know formally introduce the measure : \\\\
If $\mathcal{C}_{A}^{a}$, $\mathcal{C}_{B}^{a}$ and $\mathcal{C}_{C}^{a}$ denotes the CoA with respect to the corresponding local parties, then the volume of assistance is given by
\begin{subequations}
\begin{equation}
\label{eq:2a}
 \overline{\mathcal{C}}_3(\ket{\psi}) = \sqrt[3]{\mathcal{C}_{\scriptscriptstyle A}^{a} (\ket{\psi}) \cdot \mathcal{C}_{\scriptscriptstyle B}^{a} (\ket{\psi}) \cdot \mathcal{C}_{\scriptscriptstyle C}^{a} (\ket{\psi})}
\end{equation}
\vspace{3mm}
where $\ket{\psi}$ is any pure three qubit state. VoA is monotonous under local operation and classical communication (see Appendix \ref{Appendix 1.}). \\
This measure can also be defined for mixed three qubit states by convex roof extension:
\begin{equation}
\label{eq:2b}
    \overline{\mathcal{C}}_3 (\rho)=min \sum_i p_i \overline{\mathcal{C}}_3 (\ket{\psi_i}_{\hspace{-0.2em}\scriptscriptstyle ABC})
\end{equation}
where the minimization is over all decomposition $\{p_i,\ket{\psi_i}_{\hspace{-0.2em}\scriptscriptstyle ABC}\}$ of $\rho_{\hspace{-0.1em}\scriptscriptstyle ABC}$.\\

For a generic pure state $\ket{\psi}_{\hspace{-0.2em}\scriptscriptstyle ABC}\in \mathbb{C}^d\otimes \mathbb{C}^d\otimes \mathbb{C}^d\: (d\geqslant 3)$, the monotone GCoA is computable only for a certain class of states \cite{35}. Let $\rho = \Trace_{\scriptscriptstyle C}(\ket{\psi}_{\hspace{-0.2em}\scriptscriptstyle ABC \hspace{-0.3em}}\bra{\psi})= \sum_{k=1}^n \ket{\phi_k}\bra{\phi_k}\:(n\leqslant d^2)$ where $\ket{\phi_k}$ are subnormalized such that $\braket{\phi_k}{\phi_k}$ is the $k$-th eigenvalue of $\rho$ and $\ket{\phi_k}$'s are its eigenstates. Any other decomposition of $\rho = \sum_{l=1}^m \ket{\chi_l}\bra{\chi_l}$ is given by $\ket{\chi_l}=\sum_{k=1}^n U_{lk}^{*}\ket{\phi_k}$, for $m\geqslant n$ and $U=(U_{lk})$ being a $m\times m$ unitary operator.\\
For $d=2$, we can define the symmetric tensor of rank $2$\\
$\tau_{kk'}^\phi = \braket{\phi_k}{\Tilde{\phi_{k'}}}=det
\begin{pmatrix}
a_{11}^{k} & a_{12}^{k}\\
a_{21}^{k'} & a_{22}^{k'}
\end{pmatrix}
+det
\begin{pmatrix}
a_{11}^{k'} & a_{12}^{k'}\\
a_{21}^{k} & a_{22}^{k}
\end{pmatrix}$
where $\ket{\phi_k}=\sum_{i,j=1}^d a_{ij}^{k^{*}}\ket{ij}$ and $\ket{\Tilde{\phi_{k'}}}=\sigma_y \otimes \sigma_y \ket{\phi_{k'}^*}$.\\
For $d>2$, $\tau_{k_1 k_2 .......k_d}^\phi = \frac{d^{d/2}}{d!}\sum_\sigma det [(a_{ij}^{k_{\sigma (i)}})_{d\times d}]$ where $\sigma$ varies over all decompositions on $d$ symbols and the factor $\frac{d^{d/2}}{d!}$ has been chosen such that $\abs{\tau_{kk......k}}^{2/d}$ is the GC of the subnormalized state $\ket{\phi_k}$.\\
Hence GC of the ensemble $\rho = \sum_{k=1}^n \ket{\phi_k}\bra{\phi_k}$ is expressed as $\sum_{k=1}^n \mathcal{G}(\ket{\phi_k})=\sum_{k=1}^n \abs{\tau_{kk......k}}^{2/d}$.\\
For $d=2$, the tensor $\tau_{kk'}^\phi$ is symmetric and diagonalizable. But for $d>2$, $\tau_{k_1 k_2 .......k_d}^\phi$ is symmetric but not necessarily diagonalizable. Define the class $\mathcal{D}$ such that $\rho\in\mathcal{D}$ implies $\tau_{k_1 k_2 .......k_d}^\phi$ is a diagonalizable tensor. It is thus possible to define GCoA for any $\ket{\psi}_{\hspace{-0.2em}\scriptscriptstyle ABC}$ with $\rho_{\hspace{-0.1em}\scriptscriptstyle AB}=\Trace_{\scriptscriptstyle C}{\ket{\psi}_{\hspace{-0.2em}\scriptscriptstyle ABC\hspace{-0.3em}}\bra{\psi}}\in\mathcal{D}$ and is defined by $G_{\scriptscriptstyle C}^{a}(\ket{\psi}_{\hspace{-0.2em}\scriptscriptstyle ABC}) = \sum_{k=1}^n \lambda_{k}^{2/d}$ where $\tau_{k_1 k_2 .......k_d}^\phi = \lambda_{k_i}\prod_{j=1}^d \delta_{k_i k_j}$.\\
For any $\ket{\psi}\in \mathbb{C}^d\otimes \mathbb{C}^d\otimes \mathbb{C}^d$, $\overline{\mathcal{G}}_3 (\ket{\psi})$ is a tripartite GME measure given by
\begin{equation}
\label{eq:2c}
\overline{\mathcal{G}}_3 (\ket{\psi}) =\sqrt[3]{{\mathcal{G}_{\scriptscriptstyle A}^{a}} (\ket{\psi})\cdot {\mathcal{G}_{\scriptscriptstyle B}^{a}} (\ket{\psi})\cdot {\mathcal{G}_{\scriptscriptstyle C}^{a}}(\ket{\psi})}
\end{equation}
where $\Trace_{i} (\ket{\psi}\bra{\psi})=\rho_{jk}\in\mathcal{D}$ for all distinct $i,j,k\in\{ A,B,C\}$

This measure can also be defined for mixed three
qudit states by convex roof extension:
\begin{equation}
\label{eq:2d}
    \overline{\mathcal{G}}_3 (\rho)=min \sum_i p_i \overline{\mathcal{G}}_3 (\ket{\psi_i}_{ABC})
\end{equation}
\end{subequations}
where the minimization is over all decomposition of $\rho_{\hspace{-0.1em}\scriptscriptstyle ABC}= \sum_j q_j \ket{\psi_j}\bra{\psi_j}$.\\
Thus for certain classes of symmetric $d$-dimensional $(d\geqslant3)$ tripartite states, the measure can prove its competence. We now proceed further for extending the measure to quadripartite systems.
\subsection{Four qubit states}
Analogous to the three qubit states, it is also possible to define EoA for four qubit states. For a four qubit pure state $\ket{\psi}_{\hspace{-0.2em}\scriptscriptstyle ABCD}$, let us define
\begin{equation*}
    \mathcal{C}_{\scriptscriptstyle D}^{a}(\ket{\psi}_{\hspace{-0.2em}\scriptscriptstyle ABCD})= \max \sum_i p_i \overline{\mathcal{C}}_3 (\ket{\phi_i}_{\hspace{-0.2em}\scriptscriptstyle ABC})
\end{equation*}
where the maximization is over all decomposition $\{ p_i,\ket{\phi_i}_{\hspace{-0.2em}\scriptscriptstyle ABC}\}$ of the state $\rho_{\hspace{-0.1em}\scriptscriptstyle ABC}=\Trace_{\scriptscriptstyle D} (\ket{\psi}_{\hspace{-0.2em}\scriptscriptstyle ABCD}\bra{\psi})$, $\overline{\mathcal{C}}_3 $ denotes the volume of assistance (\ref{eq:2a}) of the three qubit state $\ket{\phi_i}_{\hspace{-0.2em}\scriptscriptstyle ABC}$ and $\mathcal{C}_{\scriptscriptstyle D}^{a}(\ket{\psi}_{\hspace{-0.2em}\scriptscriptstyle ABCD})$ denotes the entanglement of assistance with respect to party $D$. This definition can also be extended to mixed states through convex roof. We are to show that $\mathcal{C}_{\scriptscriptstyle D}^{a}(\cdot)$ is a valid four-partite monotone. Let us first start by showing that the following lemmas hold:

\textit{Lemma 1.}
    $\mathcal{C}_{\scriptscriptstyle D}^{a}(\ket{\psi}_{\hspace{-0.2em}\scriptscriptstyle ABCD})$ is invariant under $\mathcal{SL}(2,\mathbb{C})$. [see Appendix \ref{Appendix 2.}]\\

\textit{Lemma 2.}
    $\mathcal{C}_{\scriptscriptstyle D}^{a}(\ket{\psi}_{\hspace{-0.2em}\scriptscriptstyle ABCD})$ is a concave function of $\rho_{\hspace{-0.1em}\scriptscriptstyle ABC}$. [see Appendix \ref{Appendix 3.}]\\\\
We are now equipped to prove the next theorem:\\\\
\textit{Theorem 1.}
    $\mathcal{C}_{\scriptscriptstyle D}^{a}(\ket{\psi}_{\hspace{-0.2em}\scriptscriptstyle ABCD})$ is a quadripartite entanglement monotone.

\begin{proof}
    Given a pure state $\ket{\psi}_{\hspace{-0.2em}\scriptscriptstyle ABCD}$ shared among four parties Alice, Bob, Charlie and David, we will show that $\mathcal{C}_{\hspace{-0.2em}\scriptscriptstyle D}^{a}(\ket{\psi}_{\hspace{-0.2em}\scriptscriptstyle ABCD})$ does not increase on average under LOCC. Since $D$ is the assisting party, we assume that $D$ first performs measurement a given by a set of Kraus operators $\{\mathbf{M}_i\}$. Based on the outcome $i$ being sent classically to $A$ by $D$, A performs a measurement in $\{\mathbf{A}_{j}^{i}\}$ on her qubit and send the result $j$ to $B$. Upon receiving the outcome $j$, $B$ measures his system by $\{\mathbf{B}_{k}^{ij}\}$ and send the result $k$ to $C$ and then $C$ measures his system with $\{\mathbf{C}_{l}^{ijk}\}$. Finally $C$ sends the result to $D$ and $D$ performs the measurement $\{\mathbf{D}_{ijkl}^{m}\}$ depending on the outcome $l$ of $C$. This transforms the given state $\ket{\psi}_{\hspace{-0.2em}\scriptscriptstyle ABCD}$ into the mixed state $\{p_{ijklm},\ket{\phi^{ijklm}}_{\hspace{-0.2em}\scriptscriptstyle ABCD}\}$ such that
\begin{equation*}
    \ket{\phi^{ijklm}}_{\hspace{-0.25em}\scriptscriptstyle ABCD}=\frac{\mathbf{A}_{j}^{i}\otimes \mathbf{B}_{k}^{ij}\otimes \mathbf{C}_{l}^{ijk}\otimes \mathbf{D}_{ijkl}^{m} \mathbf{M}_i}{\sqrt{p_{ijklm}}} \ket{\psi}_{\hspace{-0.2em}\scriptscriptstyle ABCD}
\end{equation*}
    where $p_{ijklm}$s are the corresponding probabilities satisfying $\sum_{ijklm} p_{ijklm} = 1$.\\
 Therefore the average entanglement is given by
    \begin{equation*}
        \begin{split}
           & \sum_{ijklm} p_{ijklm} \mathcal{C}_{\scriptscriptstyle D}^{a} (\ket{\phi^{ijklm}}_{\hspace{-0.2em}\scriptscriptstyle ABCD}) \leqslant \mathcal{C}_{\scriptscriptstyle D}^{a}(\ket{\psi}_{\hspace{-0.2em}\scriptscriptstyle ABCD})
        \end{split}
    \end{equation*}
This proves that $\mathcal{C}_{\scriptscriptstyle D}^{a}(\ket{\psi}_{\hspace{-0.2em}\scriptscriptstyle ABCD})$ is a valid four qubit entanglement monotone.(see Appendix \ref{Appendix 4.} for the derivation of the inequality)
\end{proof}
Despite being a monotone, $\mathcal{C}_{\scriptscriptstyle D}^{a}(\cdot)$ fails to be a four party GME measure. This can be seen from the fact that $\mathcal{C}_{\scriptscriptstyle D}^{a}(\ket{\psi})=1$ where $\ket{\psi}=\frac{1}{\sqrt{2}}\ket{000+111}_{\hspace{-0.2em}\scriptscriptstyle ABC}\ket{0}_{\hspace{-0.2em}\scriptscriptstyle D}$ is a non-GME four qubit state. Nevertheless we can construct a four qubit GME measure by taking into account all of the four monotones with respect to different parties. For any $\ket{\psi}\in \mathbb{C}^2\otimes \mathbb{C}^2\otimes \mathbb{C}^2\otimes \mathbb{C}^2$
\begin{subequations}
\begin{equation}
\label{eq:3a}
\overline{\mathcal{C}}_4 (\ket{\psi})= \sqrt[4]{\mathcal{C}_{\scriptscriptstyle A}^{a}(\ket{\psi})\cdot \mathcal{C}_{\scriptscriptstyle B}^{a}(\ket{\psi})\cdot \mathcal{C}_{\scriptscriptstyle C}^{a}(\ket{\psi})\cdot \mathcal{C}_{\scriptscriptstyle D}^{a}(\ket{\psi})}
\end{equation}
is a quadripartite GME measure.\\
It can also be defined for mixed four qubit states through convex roof extension:
\begin{equation}
\label{eq:3b}
\overline{\mathcal{C}}_4 (\rho)=min\sum_i p_i \overline{\mathcal{C}}_4 (\ket{\psi_i}_{\hspace{-0.2em}\scriptscriptstyle ABCD})   
\end{equation}
\end{subequations}
where the minimization is over all decomposition $\{ p_i, \ket{\psi_i}_{\hspace{-0.2em}\scriptscriptstyle ABCD}\}$ of $\rho_{\hspace{-0.1em}\scriptscriptstyle ABCD}$.\\

\section{Discussion with some examples}
\label{sec4}
We now provide some examples of tripartite and four partite states and their VoA. First we provide a table of several tripartite states and their VoA. Next we show that VoA is an upper of GGM for specific classes of three qubit states. Later we show advantage of VoA over MPC \cite{48} and then test our measure for a three qubit Heisenberg model. 
\begin{center}
    \begin{tabular}{|m{1.7cm}|m{2.1cm}|}
    \hline
  \vspace{1mm}  \hspace{4mm} $\ket{GHZ}$  & \hspace{3mm} 1\\
      \hline
     \vspace{2mm}\hspace{4mm}  $\ket{W}$  & \hspace{3mm} $\frac{2}{3}$ \vspace{1mm}\\
       \hline
    \vspace{1mm}\hspace{4mm}   $\ket{\psi_{\scriptscriptstyle W}}$ & \hspace{1mm} $0.63$ \\
       \hline
    \vspace{1mm}   \hspace{4mm}   $\ket{\psi_2}$ & \hspace{1mm} $0.71$ \\
        \hline
   \vspace{1mm}     \hspace{4mm}   $\ket{\psi_3}$ & \hspace{1mm} $0.79$ \\
        \hline
       \vspace{2mm}  \hspace{4mm}   $\ket{\phi}$ & \hspace{1mm} $4\sqrt[4]{p_0 p_1 p_2 p_3}$  \vspace{1mm} \\
        \hline
       
    \end{tabular}
\end{center}
    \textit{(i)} Consider the states
    \begin{equation*}
\begin{split}
   &\ket{GHZ}=\frac{1}{\sqrt{2}}\ket{000+111} \\
 &\ket{W}=\frac{1}{\sqrt{3}}\ket{100+010+001}
    \end{split}
\end{equation*}
Then $\overline{\mathcal{C}}_3 (\ket{GHZ})=1$ and $\overline{\mathcal{C}}_3 (\ket{W})=\frac{2}{3}$. Our measure ranks the GHZ-state as more entangled than W. This verifies the fact that $\ket{GHZ}$ performs better than $\ket{W}$ in several scenarios viz. teleportation, dense coding.\\\\
\textit{(ii)} The state $\ket{\psi_{\scriptscriptstyle W}}_{\hspace{-0.2em}\scriptscriptstyle ABC}=\frac{1}{2}(\ket{100}+\ket{010}+\sqrt{2}\ket{001})_{\hspace{-0.1em}\scriptscriptstyle ABC}$ is a W-class state shared among the parties $A$, $B$ and $C$. Its VoA value $\overline{\mathcal{C}_3} (\ket{\psi{\scriptscriptstyle W}})\approx 0.63$, 
which is notably less than that of the state $\ket{W}$ ($\overline{\mathcal{C}}_3 (\ket{W}) = \frac{2}{3}$). This behavior of our measure can be explained as the lack of teleportation capability between arbitrary pairs of parties. To elaborate, the state $\ket{\psi{\scriptscriptstyle W}}$ enables perfect teleportation \cite{36} among the parties in $A|BC$ bipartition. On the other hand, from $\ket{W}$ it is possible to locally generate a Bell state using random-pair distillation \cite{52,53}, which can then be used as a viable resource for perfect teleportation. So in this context, $\ket{W}$ is more resourceful than the state $\ket{\psi_{\scriptscriptstyle W}}$. Since entanglement is inherently associated with teleportation, an ideal measure $\Lambda$ is likely to satisfy the inequality $\Lambda\left(\ket{\psi_{\scriptscriptstyle W}}\right) < \Lambda\left(\ket{W}\right)$. Additionally, it is crucial to note that the aforementioned Bell distillation from $\ket{W}$ is not possible with certainty, albeit it occurs with a high success rate. In contrast the teleportation capability of $\ket{\psi_{\scriptscriptstyle W}}$ is deterministic. Consequently, a measure that appropriately captures these nuances should not overly favor $\ket{W}$ over the state $\ket{\psi_{\scriptscriptstyle W}}$. VoA goes along with all these arguments, unlike certain previous measures. For instance, the MPC \cite{48} value for $\ket{W}$ is $\frac{2}{3}$, while it is $\frac{1}{2}$ for the state $\ket{\psi_{\scriptscriptstyle W}}$, indicating a considerably lower efficiency for $\ket{\psi_{\scriptscriptstyle W}}$ in this scenario. Such low efficiency occurs because MPC considers only one of the CoA values, unlike VoA.\\
\textit{(iii)} Consider two states $\ket{\psi_2}=\cos({\frac{\pi}{8}})\ket{000}+\sin({\frac{\pi}{8}})\ket{111}$ and $\ket{\psi_3}=\frac{1}{2}(\ket{000}+\ket{100}+\sqrt{2}\ket{111})$ mentioned in \cite{6}. It has been shown that the $C_{GME}$ measure \cite{17,37} predicts both the states as equally entangled. However, $\overline{\mathcal{C}}_3 (\ket{\psi_2})\approx 0.71 < \overline{\mathcal{C}}_3 (\ket{\psi_3})\approx 0.79$.\\\\
\textit{(iv)} The state $\ket{\phi}=\sqrt{p_0}\ket{000}+\sqrt{p_1}\ket{111}+\sqrt{p_2}\ket{222}+\sqrt{p_3}\ket{333}$ is a state of the class $\mathcal{D}$ \cite{35}. Then $\overline{\mathcal{G}}_3 (\ket{\phi})=4\sqrt[4]{p_0 p_1 p_2 p_3}=\mathcal{G}_{\scriptscriptstyle A}^{a}=\mathcal{G}_{\scriptscriptstyle B}^{a}=\mathcal{G}_{\scriptscriptstyle C}^{a}$, due to symmetry.\\\\
\textit{(v)} Next we define four highly entangled four-qubit states \cite{38} $\ket{GHZ_4}$, $\ket{W_4}$, the cluster state $\ket{C_4}$ and the Higuchi-Sudbery state \cite{39} $\ket{HS}$ such that 
\begin{equation*}
\begin{split}
       & \ket{GHZ_4} = \frac{1}{\sqrt{2}}  \ket{0000+1111}\\
        &\ket{W_4} = \frac{1}{2}  \ket{1000+0100+0010+0001}\\
        &\ket{C_4} = \frac{1}{2}  \ket{0000+0011+1100-1111}\\
        &\ket{HS} = \frac{1}{\sqrt{6}}  \bigl(\ket{0011+1100}+\omega \ket{0101+1010}\\
           & \hspace{19mm}  +\omega ^2 \ket{0110+1001}\bigr)
\end{split}
\end{equation*}
where $\omega$ denotes the cube root of unity.\\
By some elementary rearrangements of the reduced density matrices, we have $\overline{\mathcal{C}_4}(\ket{GHZ_4})=1$, $\frac{1}{2}\leqslant \overline{\mathcal{C}_4}(\ket{W_4})<1$, $\overline{\mathcal{C}_4}(\ket{C_4})=1$, $\frac{2}{3}\leqslant \overline{\mathcal{C}_4}(\ket{HS})<1$.\\
In some formerly defined measures like  \enquote{genuine multipartite concurrence}(GMC) \cite{17,37}, \enquote{geometric mean of bipartite concurrences} (GBC) \cite{12}, \enquote{concurrence fill} \cite{40} the entanglement of $\ket{C_4}$ has been shown to equal, greater and less than that of the $\ket{GHZ_4}$ state respectively. According to our measure, they are equally entangled. It is to be noted that GMC and our VoA are not equivalent as they have been shown to behave differently in example \textit{(iii)}. The equality of entanglement of the two states under VoA is due to the fact that both can be deterministically converted to tripartite GHZ state in any tripartition. Another surprising consequence of VoA is that it ranks GHZ state higher than that of the Higuchi-Sudbery state. This can be explained by the fact that $\ket{HS}$ state cannot be converted deterministically into $\ket{GHZ}$ by any local protocol. \\

\subsection{Comparison with GGM}
The generalized geometric measure (GGM) \cite{45} of a N-qubit state $\ket{\Psi^N}$ is given by \\
\[
\mathscr{G}(\ket{\Psi^N}) = 1 - \max_{\substack{\mathcal{S}_{\mathcal{A:B}}}}\{ \lambda_{\scriptscriptstyle \mathcal{A:B}}^2\}
\]
where $\lambda_{\scriptscriptstyle \mathcal{A:B}}$ is the largest Schmidt coefficient of $\ket{\Psi^N}$. The set $\mathcal{S_{A:B}}$ is the set of all arbitrary disjoint $\mathcal{A:B}$ bipartitions so that $\mathcal{A}\cup\mathcal{B}=\{1,2,3,...,N\}$.\\
Another quantity related to GGM is the localizable GGM (LGGM) \cite{45}. Let $\mathcal{P}$ be the local projective measurement on a qubit of $\ket{\Psi_N}$. It can be represented by two rank-one projective operators. The LGGM  of a N-qubit pure state is defined by
\begin{center}
    \[E_{L}^r = \sup_{\substack{\mathcal{P}}} \sum_{\substack{l=1}}^2 p^l \mathscr{G}(\ket{\Psi_{N-1}^l})\]
\end{center}
where $r\in\{1,2,...,N\}$ denotes the position of the qubit on which the measurement is performed and $p^l$ denotes the corresponding probabilities.\\
Consider the three qubit generalized GHZ (gGHZ) state $\ket{\psi_{\alpha}}=\alpha \ket{000}+ \sqrt{1-\alpha^2}\ket{111}$. Then its GGM is given by $\mathscr{G}(\ket{\psi_\alpha})=\min\{\alpha^2, 1-\alpha^2\}$. In fact, $E_L^1 =\mathscr{G}$ and due to symmetry, $E_L^1 = E_L^2= E_L^3$. But VoA of the state is given by  $\overline{\mathcal{C}}_3 (\ket{\psi_\alpha})=2\alpha\sqrt{1-\alpha^2}$. Clearly $\mathscr{G}(\ket{\psi_\alpha})<\overline{\mathcal{C}}_3 (\ket{\psi_\alpha})$.

\textit{Observation 1.-} VoA forms an upper bound of GGM and LGGM for three qubit gGHZ states.\\
It is also worth noting that unlike GGM, VoA is differentiable in the region.\\
Next let us consider the three qubit generalized W (gW) state $\ket{\psi(x_1,x_2,x_3)}=x_1\ket{000}+x_2\ket{010}+x_3\ket{001}$ such that $x_1^2+x_2^2+x_3^2=1$. It is easy to verify that VoA of the state is given by $\overline{\mathcal{C}}_3 (\ket{\psi(x_1,x_2,x_3)})=2(x_1x_2x_3)^{2/3}$. However the GGM is given by $\mathscr{G}(\ket{\psi(x_1,x_2,x_3)})=\min\{x_1^2,x_2^2,x_3^2\}$. Thus, $\overline{\mathcal{C}}_3>\mathscr{G}$ for gW states. Unlike for gGHZ states, LGGM forms a upper bound of the GGM \cite{45} for gW states. It is therefore immediate to ask if VoA also bounds the LGGM. However we find that this is not the case.

\textit{Observation 2.-} VoA forms an upper bound of GGM for three qubit gW states.\\

\subsection{Advantage over MPC} 

A novel three qubit GME measure minimum pairwise concurrence (MPC) has recently been introduced by Dong \textit{et.} \textit{al.} \cite{48}. The measure is defined by 
\begin{equation}
\label{eq:4}
\mathcal{M}_{\scriptscriptstyle ABC} = \min \{\mathcal{C}_{\hspace{-0.2em}\scriptscriptstyle A'B'}, \mathcal{C}_{\hspace{-0.2em}\scriptscriptstyle A'C'}, \mathcal{C}_{\hspace{-0.13em}\scriptscriptstyle B'C'}\}
\end{equation}
where $\mathcal{C}_{\hspace{-0.2em}\scriptscriptstyle A'B'}=\sqrt{\mathcal{C}_{\hspace{-0.2em}\scriptscriptstyle AB}^2 + \uptau_{\hspace{-0.2em}\scriptscriptstyle ABC}}$, $\mathcal{C}_{\hspace{-0.2em}\scriptscriptstyle A'C'}=\sqrt{\mathcal{C}_{\hspace{-0.2em}\scriptscriptstyle AC}^2 + \uptau_{\hspace{-0.2em}\scriptscriptstyle ABC}}$ and $\mathcal{C}_{\hspace{-0.13em}\scriptscriptstyle B'C'}=\sqrt{\mathcal{C}_{\hspace{-0.13em}\scriptscriptstyle BC}^2 + \uptau_{\hspace{-0.2em}\scriptscriptstyle ABC}}$ are called the pairwise concurrences of the respective bipartitions. $\uptau_{\scriptscriptstyle ABC}$ is the three-tangle of the corresponding three qubit pure state defined by $\uptau_{\hspace{-0.2em}\scriptscriptstyle ABC}=\mathcal{C}_{\hspace{-0.2em}\scriptscriptstyle A|BC}^2 - \mathcal{C}_{\hspace{-0.2em}\scriptscriptstyle AB}^2 - \mathcal{C}_{\hspace{-0.2em}\scriptscriptstyle AC}^2$ \cite{18}. This measure has been shown to outperform previously existing measures in many aspects. Despite being a superior GME measure, we find that there are states that MPC fails to distinguish. For instance, consider the states : 
\begin{equation*}
\begin{split}
\ket{\psi_3} &=\frac{1}{2}(\ket{000}+\ket{100}+\sqrt{2}\ket{111})\\
\ket{\psi_4} &=\frac{1}{2\sqrt{2}}(\sqrt{2}\ket{000}+\ket{100}+\ket{101}+2\ket{111})
\end{split}
\end{equation*}
We find that MPC fails to discriminate between these two states. In fact, $\mathcal{M}(\ket{\psi_3})=\frac{1}{\sqrt{2}}=\mathcal{M}(\ket{\psi_4})$. But according to VoA, $\overline{\mathcal{C}}_3(\ket{\psi_3})=\frac{1}{\sqrt[3]{2}}\approx 0.794$ and $\overline{\mathcal{C}}_3(\ket{\psi_4})=\frac{\sqrt[6]{15}}{2}\approx 0.785$. Therefore VoA predicts $\ket{\psi_3}$ to be more genuinely entangled than that of $\ket{\psi_4}$. The state $\ket{\psi_3}$ happens to be a more useful state in the sense that a maximally entangled two qubit state can be concentrated between B and C with the local assistance from A. \\
Actually this is not an isolated example. We find that there is a whole class of states that the MPC fails to discriminate. Let us define two class of pure three qubit states:
\begin{equation*}
\begin{split}
\ket{\phi_1} &= \lambda_0 \ket{000} + \lambda_1 \ket{100} + \lambda_4 \ket{111}\\
\ket{\phi_2} &=  \lambda_0 \ket{000} + \mu  \ket{100} + \lambda_2 \ket{101} + \lambda_4 \ket{111}
\end{split}
\end{equation*}
where $\lambda_0,\lambda_1,\lambda_4 \geqslant 0$ satisfying $\lambda_0^2 + \lambda_1 ^2 + \lambda_4^2 = 1$; $\mu, \lambda_2 \geqslant 0$ satisfying $\lambda_0^2  +\mu^2+ \lambda_2^2+\lambda_4 ^2 =1 $ and $\mu^2 + \lambda_2^2 = \lambda_1^2$.\\
A simple calculation shows that MPC of both $\ket{\phi_1}$ and $\ket{\phi_2}$ is given by $\mathcal{M}(\ket{\phi_1})=\mathcal{M}(\ket{\phi_2})=2\lambda_0 \lambda_4$. Nevertheless, according to VoA,
\begin{figure}[h!]
		\centering
		\includegraphics[scale=0.45]{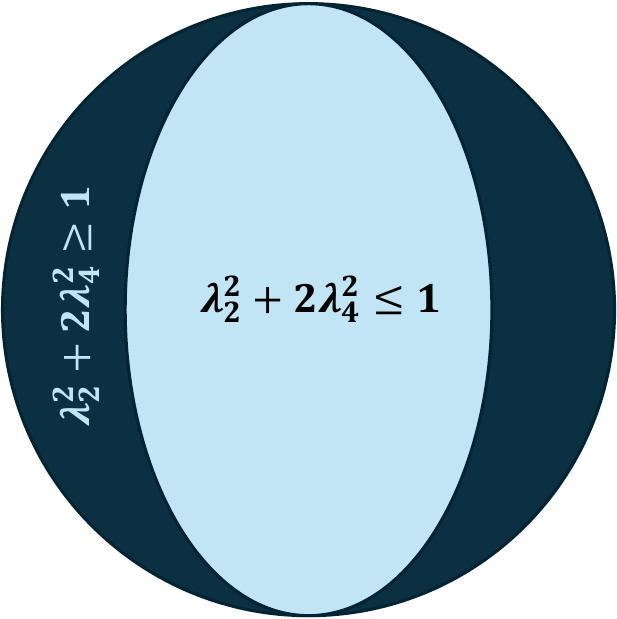}
		\caption{The inscribed ellipse denotes the area where $\overline{\mathcal{C}}_3(\ket{\phi_1}) \leqslant \overline{\mathcal{C}}_3(\ket{\phi_2})$. The outer deep blue region denotes the area where $\overline{\mathcal{C}}_3(\ket{\phi_1}) \geqslant \overline{\mathcal{C}}_3(\ket{\phi_2})$. The circle is due to the constraint $\lambda_2 ^2 + \lambda_4 ^2 \leqslant 1$.}
  \label{fig2}
	\end{figure} 
\begin{equation*}
\begin{split}
\overline{\mathcal{C}}_3(\ket{\phi_1}) &= 2\lambda_4 \sqrt[3]{\lambda_0 ^2 \sqrt{\lambda_0 ^2 + \lambda_1 ^2}}\\ 
\overline{\mathcal{C}}_3(\ket{\phi_2}) &= 2\sqrt[3]{\lambda_0 ^2 \lambda_4 ^2 \sqrt{\lambda_0 ^2 + \mu^2}\sqrt{\lambda_2 ^2 + \lambda_4 ^2}}
\end{split}
\end{equation*}
are different in general (see FIG. \ref{fig2}).\\

\subsection{VoA of a three qubit Heisenberg Model}
A three qubit Heisenberg XY model with non-uniform magnetic field is described by the Hamiltonian
\begin{equation}
\label{eq:5}
\begin{split}
  H  &= \frac{J}{2}  \sum_{i=1}^3  (\sigma_{x}^i\sigma_{x}^{i+1} + \sigma_{y}^i\sigma_{y}^{i+1})  + \frac{B_1}{2} (\sigma_{z}^1 + \sigma_{z}^{3}) \\
 &  +\frac{B_2}{2}(\sigma_{z}^2 \cos{\alpha}+ \sigma_{x}^{2}\sin{\alpha}) 
   \end{split}
\end{equation}
where J denotes the strength of the Heisenberg interaction. $B_1$ and $B_2$ are external magnetic fields with different magnitudes and $\sigma_{x,y,z}^i$ are Pauli spin operators acting on the $i$-th qubit. $J>0$ and $J<0$ correspond to the antiferromagnetic and ferromagnetic cases respectively. The angle between $B_2$ and $z$-axis of the spin is denoted by $\alpha$.\\

\begin{figure}[h!]
		\centering
		\includegraphics[scale=0.6]{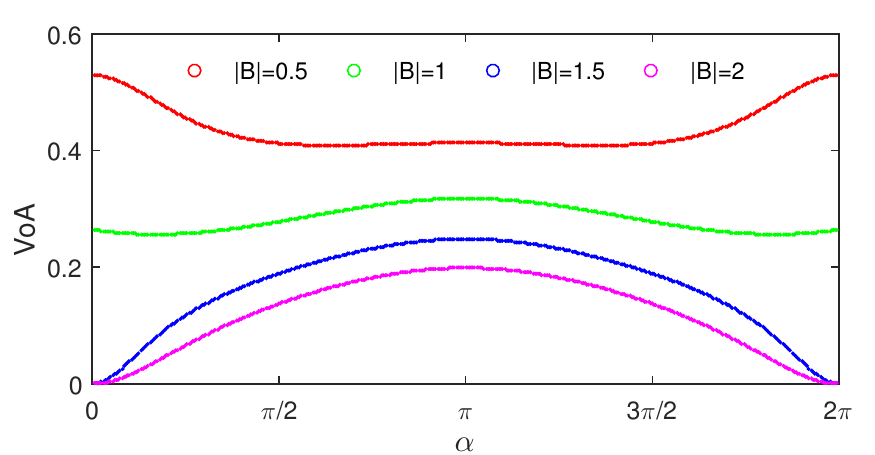}
		\caption{VoA for ground states of the model for different values of $\abs{B}$.}
  		\label{fig3}
\end{figure}

As the temperature tends to zero, the system remains in its ground state, which depends on the values of $J,B_1,B_2,\alpha$. For $\alpha \neq 0$, the eigenvalues and eigenvectors of the Hamiltonian (\ref{eq:5}) are difficult to compute. In FIG. \ref{fig3} we have measured and plotted a scattered figure \cite{qetlab} of the ground states according to VoA, by considering $J=1$ and four different values of the external magnetic field $B_1=B_2=B$. Three-tangle has been used to measure the ground state of the Hamiltonian numerically \cite{44}, for $\alpha\in (0,2\pi)$. However, as we can see, three-tangle is suboptimal to detect the genuine entanglement of three qubit ground states. The figure is symmetric at two sides about the point $\alpha = \pi$. For $\abs{B}=0.5$, the genuine entanglement is high when $\alpha$ is relatively small. But for higher values of $\abs{B}$, the scenario reverses. Higher amount of genuine entanglement is observed when $\alpha$ is near the value of $\pi$. The three-tangle has been shown to peak near the points $\alpha =\frac{\pi}{2}$ and almost vanish at $\alpha = \pi$, under identical setting. But from our figure, it is clear that despite having low three-tangle at $\alpha = \pi$, the ability of the three qubit state to localize entanglement between any two arbitrary qubits is maximal for $\abs{B}=1.5,2$. It also indicates the existence of W-type states near $\alpha = \pi$. \\

\section{Estimating Mixed States}
\label{sec5}
 
The measure can be extended to mixed states by convex roof extension. However calculating the convex roof extension for generic mixed states requires optimization over a  large number of parameters, which is beyond the scope of this work. We however provide a function to estimate the entanglement of mixed three qubit states.\\
From monogamy equality condition of pure three qubit states \cite{46}, $(\mathcal{C}_{i}^a)^2 = (\mathcal{C}_{jk})^2 + \uptau$ where $i, j, k\in \{A, B, C\}$ are all distinct, $\mathcal{C}_{jk}$ is the concurrence of the reduced density operator $\rho_{jk}$ and $\uptau$ is the three-tangle of the pure state $\ket{\psi}_{\hspace{-0.2em}\scriptscriptstyle ABC}$.\\
The three-tangle of mixed three qubit states $\rho$ is defined by 
\begin{equation*}
\uptau(\rho)=min \sum_i p_i \uptau(\ket{\psi_i})
\end{equation*}
where the minimization is over all decomposition $\{p_i, \ket{\psi_i}\}$ of $\rho$. Considering the intrinsic convexity of the concurrence over the bipartite states, we define the function
\begin{equation}
\label{eq:6}
\overline{\mathcal{C}}_3 (\rho)=\sqrt[6]{(\mathcal{C}_{\hspace{-0.2em}\scriptscriptstyle AB}^2 + \uptau(\rho))\cdot (\mathcal{C}_{\hspace{-0.2em}\scriptscriptstyle AC}^2 + \uptau(\rho))\cdot (\mathcal{C}_{\hspace{-0.2em}\scriptscriptstyle BC}^2 + \uptau(\rho))}
\end{equation}
We now try to estimate the entanglement of two particular class of mixed states.\\

\textit{Mixture of GHZ and W state.} Consider the state \\
\begin{center}
$\rho(p)=p\ket{GHZ}\bra{GHZ} + (1-p) \ket{W}\bra{W}$
\end{center}
 The three-tangle of the state is given by \cite{47}\\
\begin{center}
$\uptau\bigl( \rho(p) \bigr)=\begin{cases} 0 \hspace{6.7mm} ;\: 0\leqslant p \leqslant 0.6269... \\ g_{_I} (p) \hspace{1mm} ;\: 0.6269... \leqslant p \leqslant 0.7087... \\ g_{_{II}}(p) ;\: 0.7087... \leqslant p \leqslant 1
\end{cases}
$
\end{center}
where $g_{_I} (p) = p^2 - \frac{8\sqrt{6}}{9}\sqrt{p(1-p)^3}$ \\
and \hspace{2mm} $g_{_{II}} (p) = 1 - (1-p)(\frac{3}{2} + \frac{\sqrt{465}}{18}) $.\\
Also the concurrences $\mathcal{C}_{AB}=\mathcal{C}_{AC}=\mathcal{C}_{BC}$ are given by 
\begin{center}
$\mathcal{C}_{AB}=\begin{cases} \frac{2}{3}(1-p) - \sqrt{\frac{p(p+2)}{3}} \hspace{1mm} ;\: 0\leqslant p \leqslant 0.2918... \\ 0 \hspace{29.5mm} ;\: 0.2918... \leqslant p \leqslant 1 
\end{cases}
$
\end{center}
\begin{figure}[h!]
		\centering
		\includegraphics[width=0.45\textwidth]{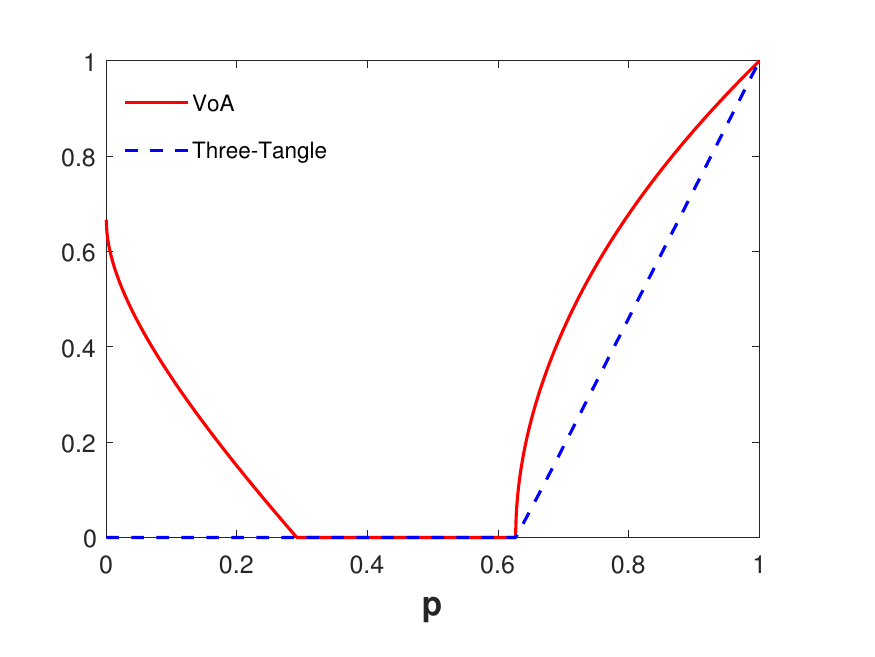}
		\caption{Comparison of $\uptau$ and $\overline{\mathcal{C}}_3$ of $\rho(p)$}
  \label{fig4}
	\end{figure} 
These two quantities together helps us to estimate the entanglement of the state $\rho(p)$. In FIG. \ref{fig4} we draw a comparison between three-tangle and VoA in estimating the genuine entanglement of $\rho(p)$.
\section{Conclusion}
\label{sec6}
The complexity of multipartite entanglement is many fold and scales exponentially with the increase in the number of parties. Its application in near term quantum technologies is multidimensional. It is therefore naive to expect a single measure to be useful in every possible scenario. The VoA provides a novel perspective within this complex domain and is meticulously designed to address a broad spectrum of scenarios, making it applicable to diverse contexts in quantum information theory. In this paper we have shown that our measure is inequivalent to some of the measures that already exist in the literature. Intriguingly our measure altercates some previously mentioned hierarchies among four qubit GME states. The measure also happens to be an upper bound of the GGM measure in specific scenarios. \\
Furthermore, we demonstrate that VoA can differentiate the varying degrees of genuine entanglement within a class of three-qubit states. In contrast, the recently proposed \enquote*{Minimum Pairwise Concurrence} (MPC) measure is unable to make these distinctions. This is particularly interesting since MPC can also be described as a function of the CoA. Notwithstanding one could argue that VoA is infact a measure that also is built upon the pairwise concurrences of a multipartite state. But this is not true since the so called monogamy equality \cite{46} does not hold beyond three qubits in general. Thus we claim that VoA is constructed from a fundamentaly different point of view. We have also provided the utility of our measure in detecting GME of the ground states of a three qubit Heisenberg XY model having non-uniform magnetic field. Later we provide a computable generalization of the VoA to mixed states.\\
VoA demonstrates an intriguing aspect in its relationship with quantum systems, highlighting its intricate dynamics. Moreover, while it poses challenges, it serves as a catalyst for further investigation and comprehension within the domain of quantum information theory. The lack of closed-form expressions for EoC \cite{33} underscores the complexity of entanglement phenomena, prompting researchers to devise novel methodologies to quantify and utilize entanglement in multipartite quantum systems, even extending beyond qubits to more generalized qudit scenarios. Although even in such cases VoA, in principle, can be defined. Nevertheless, the regularized version of the n-copy EoA : $\mathcal{E}_{\scriptscriptstyle C}^{(\infty)}(\ket{\psi}_{\hspace{-0.2em}\scriptscriptstyle ABC})=\lim_{n\to \infty}\mathcal{E}_{\scriptscriptstyle C}^{(n)}(\ket{\psi}_{\hspace{-0.2em}\scriptscriptstyle ABC})=\lim_{n\to \infty}\frac{\mathcal{E}_{\scriptscriptstyle C}(\rho_{\hspace{-0.2em}\scriptscriptstyle AB}^{\otimes n})}{n}$ (where $\rho_{\hspace{-0.1em}\scriptscriptstyle AB}=\Trace_{\scriptscriptstyle C} \ket{\psi}_{\hspace{-0.2em}\scriptscriptstyle ABC\hspace{-0.3em}}\bra{\psi}$) is a valid entanglement monotone. Thus, in asymptotic scenario VoA can be revitalized to measure the entanglement of general tripartite states.\\
 It will be interesting to see if this measure can prove its utility in other areas of quantum many body systems like quantum phase transition. Another interesting feature of multipartite entanglement measure is multipartite monogamy \cite{49,50,51}. In general, entanglement of assistance is polygamous \cite{41}. However VoA being a measure need not be \cite{42,43}. Further study is needed to resolve this problem. It will also be interesting to know if VoA can be used in scenarios other than those specified in the paper. Moreover, we would like to see if one can come up with measures that can unify some of the existing, imperfect measures to form new efficient measures that are closer to perfection.\\

\section*{ACKNOWLEDGEMENTS}
The authors I.C. and D.S. acknowledge the work as part of Quest initiatives by DST, India.The author S.B. acknowledges the support from CSIR, India. The authors A.B. and I.B. acknowledge the support from UGC, India. I.B. would also like to thank Priyabrata Char and Amit Kundu for many fruitful discussions.


\begin{thebibliography} {100}
\bibitem{Bell}  J. S. Bell, Physics Physique Fizika, 1, 195 (1964).
\bibitem{Amit1} A. Kundu, M. K. Molla, I. Chattopadhyay, and D. Sarkar, Phys. Rev.
A. 102, 052222 (2020).
\bibitem{Amit2} Amit Kundu and Debasis Sarkar, Annalen der Physik, 536(3), 2300297
(2024). 
\bibitem{Bennett} C. H. Bennett, D. P. DiVincenzo, C. A. Fuchs, T. Mor, E. Rains, P. W. Shor,
J. A. Smolin, and W. K. Wootters, Phys. Rev. A. 59, 1070 (1999).
\bibitem{secret}  D. Markham and B. C. Sanders, Phys. Rev. A 78, 042309 (2008).
\bibitem{hiding1}  B. M. Terhal, D. P. DiVincenzo and D. W. Leung, Phys. Rev. Lett. 86, 5807 (2001).
\bibitem{hiding2}  D. P. Divincenzo, D. Leung and B. M. Terhal,
 IEEE Trans. Inf. Theory 48, 580 (2002).
\bibitem{hiding3}  T. Eggeling and R. F. Werner, Phys. Rev. Lett. 89, 097905 (2002).
\bibitem{key}  G. P. Guo, C. F. Li, B. S. Shi, J. Li and G. C. Guo, Phys. Rev. A 64,
 042301 (2001).
\bibitem{Halder1} S. Halder, M. Banik, S. Agrawal, and S. Bandyopadhyay,  Phys. Rev. Lett. 122, 040403 (2019).
\bibitem{Halder2} S. Agrawal, S. Halder, and M. Banik, Phys. Rev. A. 99, 032335 (2019).
\bibitem{Atanu1}  A. Bhunia, I. Chattopadhyay, and D.
Sarkar, Quantum Information Processing, 20(1), 45 (2021).
\bibitem{Atanu2} A. Bhunia, I. Chattopadhyay, and D. Sarkar. Quantum Information Processing, 21
(5), 169 (2022).
\bibitem{Indra} I. Biswas, A. Bhunia, I. Chattopadhyay, and D. Sarkar, Physics
Letters A, 459, 128610 (2023).
\bibitem{Atanu3} A. Bhunia, I. Biswas, I. Chattopadhyay, and D. Sarkar, Journal of Physics A:
Mathematical and Theoretical, 56(36), 365303 (2023).
\bibitem{Atanu4}  A. Bhunia, S. Bera, I. Biswas, I. Chattopadhyay, and D. Sarkar, Phys. Rev. A. 109, 052211 (2024).
\bibitem{Subrata1} S. Bera, A. Bhunia, I. Biswas, I. Chattopadhyay, and D. Sarkar, Phys. Rev. A 110, 042424 (2024)
\bibitem{1} W. D\"ur, G. Vidal, and J. I. Cirac, Phys. Rev. A 62, 062314 (2000).
\bibitem{2} L. Amico, R Fazio, A. Osterloh, and V. Vedral, Rev. Mod. Phys. 80, 517 (2008), and references therein.
\bibitem{3} V. Giovannetti, S. Lloyd, and L. Maccone, Science 306, 1330 (2004).
\bibitem{4} R. Jozsa and N. Linden, Proceedings of the Royal Society
of London. Series A: Mathematical, Physical and Engineering Sciences 459, 2011 (2003).
\bibitem{5} A. Rodriguez-Blanco, A. Bermudez, M. M\"uller, and
F. Shahandeh, PRX Quantum 2, 020304 (2021).
\bibitem{6}S. Xie and J. H. Eberly, Phys. Rev. Lett. 127,
040403 (2021).
\bibitem{7} X.-F. Qian, M. A. Alonso, and J. H. Eberly, New Journal
of Physics 20, 063012 (2018).
\bibitem{8} X.-N. Zhu and S.-M. Fei, Phys. Rev. A 92, 062345 (2015).
\bibitem{9} W. K. Wootters, Phys. Rev. Lett. 80, 2245 (1998).
\bibitem{10} X. Ge, L. Liu, and S. Cheng, Phys. Rev. A 107,
032405 (2023).
\bibitem{11} Z.-X. Jin, X. Li-Jost, S.-M. Fei, and C.-F. Qiao, Phys.
Rev. A 107, 012409 (2023).
\bibitem{12} Y. Li and J. Shang, Phys. Rev. Res. 4, 023059
(2022).
\bibitem{13} K. Schwaiger, D. Sauerwein, M. Cuquet, J. I.
de Vicente, and B. Kraus, Phys. Rev. Lett. 115,
150502 (2015).
\bibitem{14} S. Puliyil, M. Banik, and M. Alimuddin, Phys. Rev. Lett.
129, 070601 (2022).
\bibitem{15} M. Huber, F. Mintert, A. Gabriel, and B. C. Hiesmayr,
Phys. Rev. Lett. 104, 210501 (2010).
\bibitem{16} A. Ghoshal, S. Choudhary, and U. Sen, All multipartite
entanglements are quantum coherences in locally distinguishable bases (2023), arXiV:2304.05249.
\bibitem{17} Z.-H. Ma, Z.-H. Chen, J.-L. Chen, C. Spengler,
A. Gabriel, and M. Huber, Phys. Rev. A 83, 062325 (2011).
\bibitem{18} V. Coffman, J. Kundu, and W. K. Wootters, Phys. Rev.
A 61, 052306 (2000).
\bibitem{19} H. Barnum and N. Linden, Journal of Physics A: Mathematical and General 34, 6787 (2001).
\bibitem{20} J. Eisert and H. J. Briegel, Phys. Rev. A 64,
022306 (2001).
\bibitem{21} M. Hein, J. Eisert, and H. J. Briegel, Phys. Rev. A
69, 062311 (2004).
\bibitem{22} D. A. Meyer and N. R. Wallach, Journal of Mathematical Physics 43, 4273 (2002), ISSN 0022-2488.
\bibitem{23} G. K. Brennen, Quantum Inf. Comput. 3, 619 (2003).
\bibitem{24}  G. Gour, D. A. Meyer, and B. C. Sanders, Phys. Rev.
A 72, 042329 (2005).
\bibitem{25} M. Choi, E. Bae, and S. Lee, Scientific Reports 13,
15013 (2023), ISSN 2045-2322.
\bibitem{26} F. Verstraete, M. Popp, and J. I. Cirac, Phys. Rev.
Lett. 92, 027901 (2004).
\bibitem{27}  M. Popp, F. Verstraete, M. A. Mart\'in-Delgado, and J. I. Cirac, Phys. Rev. A 71, 042306 (2005).
\bibitem{28} M. Horodecki, J. Oppenheim, and A. Winter, Nature 436,
673 (2005), ISSN 1476-4687.
\bibitem{29} O. Cohen, Phys. Rev. Lett. 80, 2493 (1998).
\bibitem{30} D. P. DiVincenzo, C. A. Fuchs, H. Mabuchi, J. A. Smolin,
A. Thapliyal, and A. Uhlmann,  Quantum Computing
and Quantum Communications,
(Springer Berlin Heidelberg, Berlin, Heidelberg, 1999), 247 - 257, ISBN 978-3-540-49208-5.
\bibitem{31} T. Laustsen, F. Verstraete, and S. J. van Enk, Quantum
Inf. Comput. 3, 64 (2003).
\bibitem{32} J. A. Smolin, F. Verstraete, and A. Winter, Phys. Rev.
A 72, 052317 (2005).
\bibitem{33} G. Gour and R. W. Spekkens, Phys. Rev. A 73, 062331 (2006).
\bibitem{34} G. Gour, Phys. Rev. A 74, 052307 (2006).
\bibitem{GC} I. Chattopadhyay and D. Sarkar, Quantum Information Processing 7, 243 (2008).
\bibitem{35} G. Gour, Phys. Rev. A 72, 042318 (2005).
\bibitem{36} P. Agrawal and A. Pati, Phys. Rev. A 74, 062320 (2006).
\bibitem{37} S. M. Hashemi Rafsanjani, M. Huber, C. J. Broadbent,
and J. H. Eberly, Phys. Rev. A 86, 062303 (2012).
\bibitem{38} A. Mishra, S. Mahanti, A. K. Roy, and P. K. Panigrahi, Physics Open 20, 100230 (2024).
\bibitem{39} A. Higuchi and A. Sudbery, Physics Letters A 273, 213 (2000).
\bibitem{40}  S. Xie, D. Younis, Y. Mei, and J. H. Eberly, Unraveling the mysteries of multipartite entanglement: A journey through geometry (2023), arXiV:2304.03281.
\bibitem{41} Y. Guo, Quantum Information Processing 17, 222
(2018).
\bibitem{42}  Y. Guo and L. Zhang, Phys. Rev. A 101, 032301 (2020).
\bibitem{43} C. Lancien, S. Di Martino, M. Huber, M. Piani, G. Adesso, and A. Winter, Phys. Rev. Lett. 117, 060501 (2016).
\bibitem{44} Ren Jie and Zhu Shi-Qun, Communications in Theoretical Physics, 46(6) 969, dec 2006.
\bibitem{45} D. Sadhukhan, S. S. Roy, A. K. Pal, D. Rakshit, A. Sen(De), and U. Sen, Phys. Rev. A 95, 022301 (2017) and references therein.
\bibitem{qetlab} Nathaniel Johnston. QETLAB: A MATLAB toolbox for quantum entanglement, version 0.9. https://qetlab.com, January 12, 2016.
\bibitem{46} Chang-shui Yu and He-shan Song, Phys. Rev. A 76, 022324 (2007).
\bibitem{47} R. Lohmayer, A. Osterloh, J. Siewert, and A. Uhlmann, Phys. Rev. Lett. 97, 260502 (2006).
\bibitem{48} Dong-Dong Dong , Li-Juan Li, Xue-Ke Song, Liu Ye, and Dong Wang, Phys. Rev. A 110, 032420 (2024).
\bibitem{49} M. F. Cornelio, Phys. Rev. A 87, 032330 (2013).
\bibitem{50} Y. Guo and L. Zhang, Phys. Rev. A 101, 032301 (2020).
\bibitem{51} P. Char, D. Chakraborty, P. K. Dey, A. Sen, A. Bhar, I. Chattopadhyay and D. Sarkar, Multipartite Monogamy of Entanglement for Three Qubit States, arXiv:2409.00865.
\bibitem{52} B. Fortescue and H.-K. Lo, Phys. Rev. Lett. 98, 260501 (2007).
\bibitem{53} E. Chitambar, W. Cui, and H.-K. Lo, Phys. Rev. Lett. 108, 240504 (2012).



\end{thebibliography}

\appendix
\section*{Appendix}
\subsection{LOCC monotonicity of VoA}
\label{Appendix 1.}
\textbf{H\"older's Inequality}: Let $\lambda_1,\lambda_2,...,\lambda_n\in \mathbb{R}^+$ such that $\sum_{i=1}^n \lambda_i =1$ and $\{ a_i\}_{i=1}^l, \{ b_i\}_{i=1}^l,..., \{ n_i\}_{i=1}^l$ are sets of positive reals. Then \\
\begin{center}
    $(\sum_{i=1}^l a_i)^{\lambda_1} (\sum_{i=1}^l b_i)^{\lambda_2} ... (\sum_{i=1}^l n_i)^{\lambda_n} \geqslant \sum_{i=1}^l a_{i}^{\lambda_1} b_{i}^{\lambda_2} ... n_{i}^{\lambda_n}$
\end{center}
$\bullet$
The LOCC monotonicity of the measure can be proven by applying H\"older's Inequality. Let $\rho$ be a tripartite state and after passing through some LOCC channel, it converts into the ensemble $\{ p_k, \rho_k\}$ where $k=1,2,.....,l$. Consider a set of entanglement monotones $\Lambda_i$ satisfying $\Lambda_i (\rho) \geqslant \sum_k p_k \Lambda_k (\rho_k) \geqslant 0$ where $i=1,2,...,n$.\\
\begin{equation}
    \begin{split}
        \label{A_eq1}
       & [\Lambda_1 (\rho) \Lambda_2 (\rho)...... \Lambda_n (\rho)] \\
       \geqslant & \{\sum_k p_k \Lambda_1 (\rho_k)\}^{1/n} \{\sum_k p_k \Lambda_2 (\rho_k)\}^{1/n} ... \{\sum_k p_k \Lambda_n (\rho_k)\}^{1/n}
    \end{split}
\end{equation}
Let $\lambda_i = \frac{1}{n} \hspace{1mm} \forall i$ and $a_i = p_i \Lambda_1 (\rho_i), b_i = p_i \Lambda_2 (\rho_i),..., n_i = p_i \Lambda_n (\rho_i)$.\\
\begin{equation}
    \begin{split}
    \label{A_eq2}
       & \{\sum_k p_k \Lambda_1 (\rho_k)\}^{1/n} \{\sum_k p_k \Lambda_2 (\rho_k)\}^{1/n} ... \{\sum_k p_k \Lambda_n (\rho_k)\}^{1/n} \\
       \geqslant & \sum_{k} \{ p_k \Lambda_1 (\rho_k)\}^{1/n} \{ p_k \Lambda_2 (\rho_k)\}^{1/n} ... \{ p_k \Lambda_n (\rho_k)\}^{1/n}\\  
       = & \sum_k p_k [\Lambda_1 (\rho_k)\Lambda_2 (\rho_k)......\Lambda_n (\rho_k)]^{1/n}
    \end{split}
\end{equation}
From \ref{A_eq1} and \ref{A_eq2}, we have $E(\rho)\geqslant \sum_k p_k E(\rho_k)$ where $E({\rho})=\sqrt[n]{\Lambda_1 (\rho) \Lambda_2 (\rho) ...... \Lambda_n (\rho)}$ implies that $E(\cdot)$ is a tripartite entanglement monotone.\\
\subsection{Proof of Lemma 1.}
\label{Appendix 2.}
\textit{Lemma 1.}\:
    $\mathcal{C}_{D}^{a}(\ket{\psi}_{ABCD})$ is invariant under $\mathcal{SL}(2,\mathbb{C})$.
\begin{proof}
    Let $\mathbf{A}, \mathbf{B}, \mathbf{C}\in \mathcal{SL}(2,\mathbb{C})$ denote the operations on respective subsystems of the state $\ket{\psi}$. Without loss of generality, we assume that only the operation $\mathbf{A}$ is performed on the subsystem A. By definition of EoA for four qubit systems defined above, $\mathcal{C}_{D}^{a}(\ket{\psi}_{ABCD})$ is a function of $\rho_{ABC}$ only. Let us define $\Tilde{\rho}_{ABC}$ such that\\
  \begin{center}
        $\Tilde{\rho}_{ABC}=(\mathbf{A}\otimes \mathbf{I}\otimes \mathbf{I})\rho_{ABC}(\mathbf{A}^{\dagger}\otimes \mathbf{I}\otimes \mathbf{I})$
    \end{center}
    and $\{ p_i,\ket{\phi_i}_{ABC}\}$ be the optimal decomposition of $\Tilde{\rho}_{ABC}$ such that $\mathcal{C}_{D}^{a}(\Tilde{\rho}_{ABC})=\sum_i p_i \overline{\mathcal{C}}_3 (\ket{\phi_i}_{ABC})$. Since $\mathbf{A}\in \mathcal{SL}(2,\mathbb{C})$, $\mathbf{A}^{-1}$ exists and hence,\\
\begin{equation}
        \begin{split}
        \label{A_eq3}
            \rho_{ABC}&=(\mathbf{A}^{-1}\otimes \mathbf{I}\otimes \mathbf{I})\Tilde{\rho}_{ABC}(\mathbf{A}^{-1\dagger}\otimes \mathbf{I}\otimes \mathbf{I})\\
            \Rightarrow \mathcal{C}_{D}^{a}(\rho_{ABC}) & \geqslant \sum_i p_i \overline{\mathcal{C}}_3 (\mathbf{A}^{-1}\ket{\phi_i}_{ABC}\bra{\phi_i}\mathbf{A}^{-1\dagger})\\      
             & =\sum_i p_i \: det(\mathbf{A}^{-1}) \overline{\mathcal{C}}_3 (\ket{\phi_i}_{ABC})\\
             & =\sum_i p_i \overline{\mathcal{C}}_3 (\ket{\phi_i}_{ABC})\\
             & = \mathcal{C}_{D}^{a}(\Tilde{\rho}_{ABC})
        \end{split}
    \end{equation}\\
 Conversely, let $\{ q_j,\ket{\chi_j}_{ABC}\}$ be the optimal decomposition of $\rho_{ABC}$ such that $\mathcal{C}_{D}^{a}(\rho_{ABC})=\sum_j q_j \overline{\mathcal{C}}_3 (\ket{\chi_j}_{ABC})$. This implies\\
    \begin{equation}
        \begin{split}
        \label{A_eq4}
            \mathcal{C}_{D}^{a}(\Tilde{\rho}_{ABC}) & = \mathcal{C}_{D}^{a}\bigl((\mathbf{A}\otimes \mathbf{I}\otimes \mathbf{I})\rho_{ABC}(\mathbf{A}^{\dagger}\otimes \mathbf{I}\otimes \mathbf{I})\bigl)\\
            & \geqslant \sum_j q_j \overline{\mathcal{C}}_3 (\mathbf{A}\ket{\chi_j}_{ABC}\bra{\chi_j}\mathbf{A}^{\dagger})\\
            & = \sum_j q_j \: det(\mathbf{A}) \overline{\mathcal{C}}_3 (\ket{\chi_j}_{ABC})\\
            & = \sum_j q_j \overline{\mathcal{C}}_3 (\ket{\chi_j}_{ABC})\\
            & = \mathcal{C}_{D}^{a}(\rho_{ABC})
        \end{split}
    \end{equation}\\
    Combining two equations (\ref{A_eq3}) and (\ref{A_eq4}), our statement is thus proved.
\end{proof}
\subsection{Proof of Lemma 2.}
\label{Appendix 3.}
\textit{Lemma 2.}\:
    $\mathcal{C}_{D}^{a}(\ket{\psi}_{ABCD})$ is a concave function of $\rho_{ABC}$.
\begin{proof}
   The proof is quite straightforward. By definition,\\
    \begin{equation*}
    \begin{split}
         \mathcal{C}_{D}^{a}(\ket{\psi}_{ABCD}) & = max_{\{ p_i, \ket{\phi_i}_{ABC}\}} \sum_i p_i \overline{\mathcal{C}}_3 (\ket{\phi_i}_{ABC})\\
         & =\sum_i p_i \overline{\mathcal{C}}_3 (\ket{\phi_i}_{ABC})\\
    \end{split}
    \end{equation*}\\
    \emph{i.e.}, we have assumed $\{ p_i, \ket{\phi_i}_{ABC}\}$ to be the optimal decomposition, with $p_i$ being the probabilities.\\
    Define two three qubit states $\sigma_{ABC}$, $\tau_{ABC}$ such that $\rho_{ABC} = \lambda \sigma_{ABC} + (1-\lambda)\tau_{ABC}$ with $0<\lambda<1$.\\
    We also assume $\{ q_j, \ket{\chi_j}_{ABC}\}$ and $\{ r_k, \ket{\eta_k}_{ABC}\}$ be the optimal decompositions of $\sigma_{ABC}$ and $\tau_{ABC}$ respectively, where $q_j\geqslant 0, r_k\geqslant 0$ are probabilities satisfying $\sum_j q_j =1, \sum_k r_k =1$. Then,\\
    \begin{equation*}
        \begin{split}
            & \lambda  \mathcal{C}_{D}^{a}(\sigma_{ABC}) + (1-\lambda)\mathcal{C}_{D}^{a}(\tau_{ABC})\\
            & =\lambda \sum_j q_j \overline{\mathcal{C}}_3 (\ket{\chi_i}_{ABC}) + (1-\lambda)\sum_k r_k \overline{\mathcal{C}}_3 (\ket{\eta_k}_{ABC})\\
            & \leqslant \sum_i p_i \overline{\mathcal{C}}_3 (\ket{\phi_i}_{ABC})\\
            & = \mathcal{C}_{D}^{a}(\rho_{ABC})
        \end{split}
    \end{equation*}
\end{proof}
\subsection{Proof of Theorem 1. : Deriavtion of the Inequality}
\label{Appendix 4.}
    \begin{equation*}
        \begin{split}
           & \sum_{ijklm} p_{ijklm} \mathcal{C}_{D}^{a} (\ket{\phi^{ijklm}}_{\scriptscriptstyle ABCD})\\
           & =\sum_{ijklm} \abs{det(\mathbf{A}_{j}^{i})}\abs{det(\mathbf{B}_{k}^{ij})}\abs{det(\mathbf{C}_{l}^{ijk})} \mathcal{C}_{D}^{a}(\mathbf{D}_{m}^{ijkl} \mathbf{M}_i \ket{\psi}_{ABCD})\\
            & =\sum_{ijklm} T_{ijkl} \mathcal{C}_{D}^{a}\bigl( \Trace_D \mathbf{D}_{m}^{ijkl} \mathbf{M}_i \ket{\psi}_{ABCD}\bra{\psi} \mathbf{M}_{i}^\dagger (\mathbf{D}_{m}^{ijkl})^\dagger\bigl)\\
            & \leqslant \sum_{ijkl} T_{ijkl} \mathcal{C}_{D}^{a}\bigl( \Trace_D \mathbf{M}_i \ket{\psi}_{ABCD}\bra{\psi} \mathbf{M}_{i}^\dagger \bigl)\\
            & \leqslant \sum_i  \mathcal{C}_{D}^{a}\bigl( \Trace_D \mathbf{M}_i \ket{\psi}_{ABCD}\bra{\psi} \mathbf{M}_{i}^\dagger \bigl)\\
            & \leqslant  \mathcal{C}_{D}^{a}(\rho_{ABC})\\
            &  =\mathcal{C}_{D}^{a}(\ket{\psi}_{ABCD})
        \end{split}
    \end{equation*}
where $T_{ijkl}=\abs{det(\mathbf{A}_{j}^{i})}\abs{det(\mathbf{B}_{k}^{ij})}\abs{det(\mathbf{C}_{l}^{ijk})}$. \\
The first and last inequality comes from the fact that $\mathcal{C}_{D}^{a}(\cdot)$ is concave in $\rho_{ABC}$. The second inequality is due to the Arithmetic Mean - Geometric Mean inequality such that $\sum_j \abs{det(\mathbf{A}_{j}^i)}\leqslant \frac{1}{2} \sum_j \mathbf{A}_{j}^{i\dagger}\mathbf{A}_{j}^i =1$ and similarly for $\mathbf{B}_{k}^{ij}$, $\mathbf{C}_{l}^{ijk}$. 
\end{document}